\documentclass[11pt]{amsart}
\usepackage{geometry}                
\usepackage[parfill]{parskip}    
\usepackage{graphicx}
\usepackage{amssymb}
\usepackage{epstopdf}
\usepackage{url}
\usepackage{graphicx}

\def\keywords{\vspace{.5em}
{\textit{Keywords}:\,\relax%
}}
\def\endkeywords{\par}

\begin{document}
%
\title{Designing anti-cancer drugs and directing anti-cancer drug therapy }
%
%
%
%

\author{Elinor Velasquez, Jorge Soto-Andrade, Ben Bongalon}
\date{Draft 1.0 (December, 2013)}

%
%

\markboth{}%
{Shell \MakeLowercase{\textit{}}: A computer visualization tool for designing anti-cancer drugs and directing anti-cancer drug therapy }
%


\begin{abstract}
 A prototype for a web application was designed and implemented as a guide to be used by clinicians when designing the best drug therapy for a specific cancer patient, given biological data derived from the patientÕs tumor tissue biopsy. A representation of the patientÕs metabolic pathways is displayed as a graph in the application, with nodes as substrates and products and edges as enzymes. The top metabolically active sub-paths in the pathway, ranked using an algorithm based on both the patientÕs biological data and the graph topology, are also displayed and can be individually highlighted to examine potential enzymatic sites to be disrupted by a drug Ð these sites serve as a guide for designing the patientÕs specific drug therapy. Displayed next to each sub-path is the sub-path score used to decide its rank, as well as the predicted patient survival time to indicate how effective that specific drug will be in alleviating that patient's cancer. Future work includes an animation component to track the patientÕs progress and re-design the drug therapy as needed. Corresponding author's academic email: elinor@soe.ucsc.edu.
  \end{abstract}


\maketitle


%

\keywords
{\bf{Keywords:}} visualization, metabolic pathway, enzyme-inhibiting drugs
\endkeywords

\hfill 
\section{{\bf{INTRODUCTION}}}
One non-invasive method for removing a patientÕs cancer is the use of chemotherapy or application of cytotoxic drugs to destroy the cancer cells. The currently used cytotoxic drugs were designed to damage the DNA in a cancer cell, but have produced mixed results for destroying the cancer cells and restoring good health to the patient. Cancer is believed to arise from a threshold number of gene mutations occurring in one or more cells. An alternative viewpoint, not mutually exclusive of the gene mutation viewpoint, is that cancer is a metabolic disease. While metabolic pathway activity is similar across healthy patients or humans without cancer, metabolic pathway activity varies across cancer patients. Thus, each patientÕs cancer is unique to that patient, giving rise to a need for a personalized medicine approach to ease cancer in that patient. Viewing cancer as a metabolic disease permits the possibility of designing novel drugs, which disrupt metabolic pathways in cancer cells, rapidly causing their death and quickly restoring good health to the patient. 

The computer visualization website that is presented here is intended to contribute to alleviating cancer in a patient by providing the clinician with a computational application that 1) uses the clinical and biological data of their patient to find the best drug for disrupting the patientÕs cancer cellular metabolic pathways, 2) predicts a measure of efficacy of this drug for that patient, and 3) tracks the patientÕs progress while undergoing the drug therapy and use the previous features of the tool to make adjustments to the drug therapy as needed. 

Current visualization tools of metabolic pathways typically have the following features in common: Data from many patient samples are visualized instead of samples from a single patient. Data is presented in a raw state. It is up to the clinician to decide which node or path in the pathway to disrupt. In other words, typically nodes or paths are not ranked in order of importance to cancer cellular activity, so the user must independently propose the best drug for path disruption. Also, there is no metric to estimate drug efficacy if a certain drug is applied to a specific patient. Lastly, no animation of a patientÕs metabolic activity yet exists to track a patientÕs progress during drug therapy and vary the type of drug used as needed. 

\section{{\bf{METHODS AND MATERIALS}}}

Here are current features of the prototype website: It accepts processed (non-raw) data resulting from computations involving prior cancer, other biological studies, and the patientÕs data. The processed data is then displayed as edge scores to show the most metabolically active edges in the metabolic pathways graph, as sub-path scores to show ranking of sub-paths in the graph, and as patient survival time associated with each knocked out sub-path or drug.

Future features of the prototype website will permit clinicians to upload a single patientÕs raw data and run the necessary algorithms required to compute the edge scores for the graph, compute the scores and ranking of the sub-paths in the graph, compute the predicted patient survival times as a result of knocking out a sub-path using a drug, track a patientÕs progress during drug therapy, and display score uncertainty. 

The current visualization displays one metabolic pathway, the Kreb cycle, depicted as a graph with oriented edges and nodes. A node is labeled by the key substrate or product for a given reaction. Each edge that connects two nodes has two scores displayed, namely the enzymeÕs European Commission (EC) number and a score based on prior cancer and other biological studies, and the patientÕs tumor data, specifically, a patientÕs microarray mRNA expression data produced via the patientÕs tumor tissue as well as data produced via healthy tissue, with that healthy tissue from the patient if the patient has a primary tumor or from tissues of healthy humans if the patient has a metastasized tumor. The current visualization also displays a list of the top sub-paths, ranked by metabolic activity score and displayed with predicted patient survival time post-drug therapy, tabs for uploading data, selecting pathways, animation, and help. The visualization additionally permits the clinician to select a sub-path in the ranked list given on the left hand side of the metabolic pathway and highlight that sub-path as a single colored connected set of edges and nodes. 

Hovering over an edge in the prototype visualization displays a box of annotation detailing the name of the associated reaction, the Recon 2 abbreviation of the reactionÕs name, extracted from Thiele et al. (2013) supplementary files, the reaction formula in the mitochondria, the enzyme EC number, the enzymeÕs complete name, and the computed metabolic activity score for that edge. 
Materials and methods

\subsection{Edge weight describes metabolic activity}
The edge weight in a metabolic pathway graph extends a conception of gene activity described by Karr et al. (2012). Karr et al. construct an equation to estimate the probability of transcribing a gene. A score for gene transcription is derived from the rate of change of the concentration of a specified RNA and is proportional to the gene expression value multiplied by the sum of the geneÕs half-life and the cell cycle time. The gene activity is obtained from dividing this gene transcription score by the sum of all gene transcription scores. So the edge weight is a probability. 

The edge weight in a metabolic pathway graph estimates the likelihood of a certain amount of metabolic activity of an enzyme in the mitochondria during a snapshot of the reaction. The snapshot is the biological information collected at the time of a patientÕs cancer tissue biopsy or healthy tissue biopsy. The numerator of the edge weight is defined as the metabolic activity score and equals the reaction flux multiplied by the enzymeÕs half-life (obtained by mining the literature), the sum (over relevant genes) of 2x (with x the normalized microarray gene expression value obtained from the biological data derived from either the patientÕs tumor and healthy tissue biopsies) multiplied by the geneÕs half-life (obtained by mining the literature). A model for each metabolic pathway is constructed from the Recon 2 human metabolic pathways information (Thiele et al., 2013). Then each reactionÕs steady state flux is computed by applying the Copasi biochemical network simulator software to the model (www.copasi.org/tiki-view\_articles.php), and mining the literature to obtain values for the initial conditions and kinetic constants needed for the model. The edge weight is the probability obtained from dividing this edge metabolic activity score by the sum of all edge metabolic activity scores. Preferred additional data used to compute the edge weight can include biological data derived from the cancer patient, involving protein array values to estimate an enzymeÕs translation score (that multiples the enzymeÕs half-life) and C13 data values for estimating the flux of the metabolic reactions.

\subsection{Ranked sub-paths guide clinician when designing drug therapy for patient}
An algorithm was developed for ranking sub-paths using edge weights and the topology of the graph. The algorithm is an adaptation of DijkstraÕs algorithm for computing the most likely path configuration between two nodes A and B. The node A represents the substrates for a metabolic reaction, the node B represents the products of that metabolic reaction, and the edge between node A and B represents the enzyme used to catalyze the reaction. The algorithm inputs the edge weights and outputs a list of sub-paths, ranked from the sub-path with highest metabolic activity to the sub-path with least metabolic activity.

DijkstraÕs theorem finds the shortest path between two given nodes in a directed graph with weighted edges. The set of all nodes is partitioned into the subset of settled nodes, SN, whose shortest path distances from the source s are to be determined and the subset of unsettled nodes, USN. Initially, SN is empty. While USN is not empty, the node in USN with the minimal path distance to s is extracted, denoted n, and is added to SN. Its nearest neighbors are relaxed: Let a given neighbor be denoted by nb. If the (direct) path distance between s and nb, denoted d[s, nb], is greater than the (indirect) path between s and nb via n, then d[s, nb] is updated to equal the indirect path distance which equals the sum of d[s, n] and the weight of edge e, where e has initial node n and terminal node nb. The predecessor of nb is then updated to be n and the while loop repeats.

See Figure 1 for an illustration of DijkstraÕs theorem for the triangle of nodes s, nb and n. In this figure, the direct path distance between s and nb is 7 and the indirect path distance equals 2 + 3 = 6. Thus, the indirect path distance is the shortest distance between node s and node nb. 
  
\begin{figure}[htb]
       \center{ \includegraphics[width=5cm]{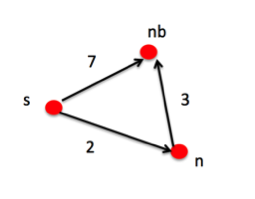} }
        \caption{An illustration of DijkstraÕs algorithm in which the indirect path is shorter than the direct path. } 
\end{figure}

Consider now a directed graph constructed for illustrating metabolic pathways. In this adaptation of DijkstraÕs algorithm, the metabolites are the nodes and the enzymes are the edges of the graph. For a given metabolic reaction, there are one or more substrates, one or more products and an enzyme. The substrates in a given metabolic reaction are all represented by one node and the associated products are all represented by another node. The enzyme for the reaction is represented by an edge between these nodes. An edge weight equals the value of the probability distribution of metabolic activity evaluated for a given reaction and for a given enzyme in the cell mitochondria. 

The implementation of the adaptation of DijkstraÕs algorithm uses four data structures. Implemented as a hash map, CurrProb is an evolving list of nodes n and their source path probabilities, P[s, n], initialized to zero for all nodes n without a direct path to the source node s and otherwise to a nonzero value such as the edge weight. CurrProb is updated as each nearest neighbor is relaxed; see below for details. The hash map Predecessor is an evolving list of nodes and their path predecessors, and is updated as each nearest neighbor is relaxed; see below for details. SN is the set of settled nodes in which the most likely path has been found from the source node to the settled node. USN is the set of unsettled nodes and equals the set of all nodes minus the set of settled nodes. Both SN and USN are implemented as priority queues and are reverse ordered according to each nodeÕs source path probability, P[s, n], with s the source node and n the node; the maximum probability is at the top of the queue. 

The input to the algorithm is a node pair (A, B). The most likely sub-path between A and B and its sub-path score, P[A, B], are output by the algorithm. The output is described by a sequence of intermediate node pairs (a, b) (the Predecessor hashmap) and P[A, B] = (1/N)· P[a, b], with N the number of node pairs (a, b). An intermediate node pair equals the initial node a and the terminal node b. P[a, b] equals the maximum edge weight over all the edges that connect node a to node b and have Hamming distance d(a, b) = 1. A wrapper is used to iterate the algorithm over all nodes A and B in the graph and the wrapper output is a list of the top sub-paths in the graph, ranked by the sub-path scores. 

The algorithm proceeds as follows. CurrProb is initialized as previously described. The source node, A, is fixed and the target node, B, is fixed. A is added to the priority queue SN and the probability of each neighbor n of A, P[A, n], is computed and the neighbor with the maximum P[A, n] is added to the priority queue USN. While USN is not empty, extract the top queued node from USN; call it ntop. Add ntop to the priority queue SN, then relax the neighbors of ntop: If neighbor nb belongs to SN, ignore this neighbor. Otherwise, check if the probability P[A, nb] is less than P[A, ntop] times P[ntop,nb]. If so, set P[A, nb] equal to P[A, ntop] times P[ntop, nb] and update both the CurrProb hash map and the USN priority queue (which is reverse ordered according to P[A, nUSN], for all nodes nUSN in the USN). The predecessor of nb is set to be ntop and the Predecessor hash map is updated. After the while loop exits, the path configuration is described by the elements in Predecessor. Lastly, the node A, the node B, the elements of Predecessor, and P[A, B] are presented as output. 

\subsection{Metric predicts efficacy of drug therapy pre-treatment}
A metric was developed to estimate the efficacy of a proposed drug therapy for a cancer patient. Current implementation of the visualization permits the clinician to inspect the ranked list of sub-paths and select one or more sub-paths to knock out. Note that the algorithm ranking sub-paths assumes that at least one reaction is involved in a sub-path. For the initial description of the metric, suppose a sub-path has only one reaction: A reaction usually has both upstream and downstream reactions to it. If a reaction is knocked out, the downstream reactions will all be knocked out as well as some of the upstream reactions if the primary reaction is reversible. The current implementation of the metric considers only downstream reactions. 

The efficacy of a proposed drug is estimated by predicting survival time (lifespan) of a cancer patient if the drug is used in the patientÕs anti-cancer drug therapy. The metric is defined as a simple linear regression, Y = · mX + b, with Y the patientÕs predicted survival time (post-drug therapy), X the edge weight for a given edge in the metabolic pathways graph, b the Y-intercept, and m the slope. The values m and b are the solution obtained by minimizing Q(m,b), with Q(m,b) = · (Yi Ð b Ð mXi)2, with the sum over all pairs (Yi, Xi): 

The pairs (Yi, Xi) are taken from prior cancer studies that contain the following data types: survival times of cancer patient who have undergone a specific enzyme-inhibitor drug therapy, and microarray gene expression values derived from patientsÕ cancer tissue biopsies and additionally healthy tissue biopsies from either the patient or non-patients (in the case of metastasized cancer). Preferred additional data used to compute the edge weight can include biological data derived from the cancer patient, involving protein array values to estimate an enzymeÕs translation score (that multiples the enzymeÕs half-life) and C13 data values for estimating the flux of the metabolic reactions.

The sum in the simple linear regression is taken over all the reactions that the enzyme catalyzes, and all reactions downstream of these disrupted reactions. 

\section{{\bf{IMPLEMENTATION}}}

The visualization component uses the GraphViz tool, http://graphviz.org/, to generate metabolic graphs. Using GraphViz brings two key benefits. First the metabolic pathways can be described at a high level, and the tool automatically takes care of laying out the graph. Secondly, GraphViz can generate the graphs in SVG (Scalable Vector Graphics) format. This allows the graph to be arbitrarily resized without pixelation. 

After the SVG graph is created, the D3.js tool, http://d3js.org/, is used to add annotations, highlight pathways and manipulate the image. D3.js is a JavaScript library for manipulating documents based on data, including SVG. The entire application is hosted on Google App Engine, https://developers.google.com/appengine/. 

Figure 2 displays the current implementation of the visualization.

 \begin{figure}[htb]
        \center{\includegraphics[width=15cm]{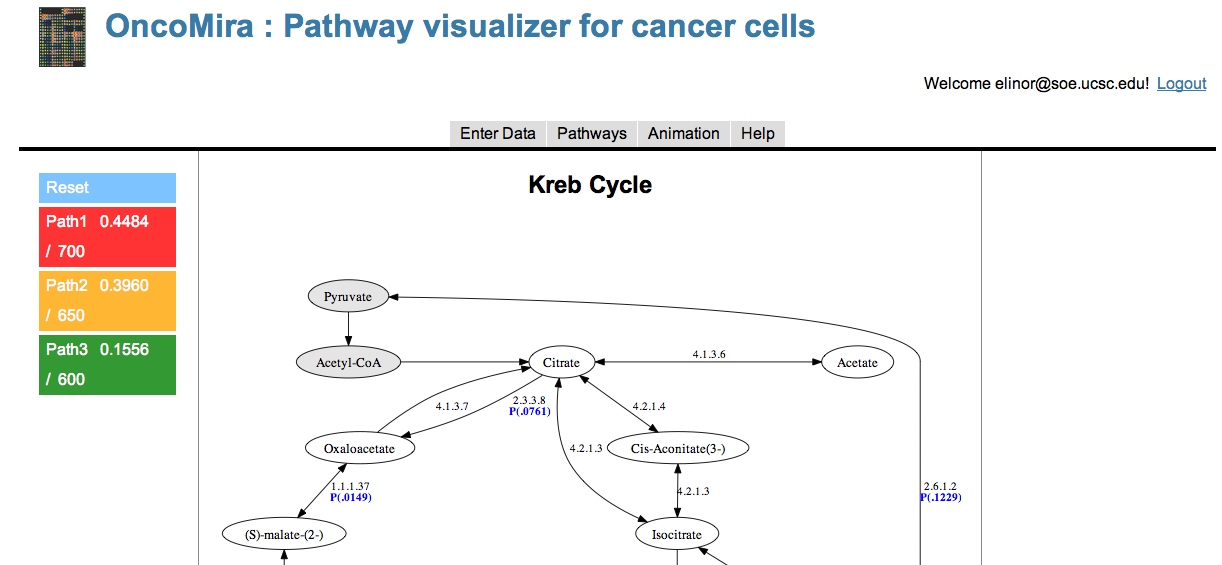} }
        \caption{OncoMira: Pathway visualizer for cancer cells application displaying the beginning of the Kreb cycle. The top three ranked sub-paths, their path scores (the listed numbers are temporary placeholders), and the patientÕs predicted survival time (the listed numbers are temporary placeholders) if a drug knocks out that sub-path are displayed on the left hand side of the screen. The blue bold number on most edges represents the edge weight. The lack of a blue bold number means that the edge does not have either patient derived or literature derived data to complete the calculation for the edge weight; the missing information has to be estimated. This information and the resulting uncertainty in the edge weight because of this lack of information will be displayed in future implementations. The black number represents the European Commission enzyme identification number for an edge. Selecting a specific sub-path colors the relevant nodes and edges on the graph. } 
\end{figure}

\section{{\bf{CASE STUDIES}}}

The metabolic pathway that is currently displayed in the visualization is the citric acid cycle (TCA or Kreb cycle) in a cellÕs mitochondria, used to completely oxidize glucose, producing energy that is carried by NAD and FADH to the electron transport chain so ATP can be outputted. Mutations in genes associated with the TCA cycle enzyme, isocitrate dehydrogenase, have already been linked to tumorogenesis [Parsens, et al., 2008], [Yan, et al., 2009]. Also, the TCA cycle enzyme, fumarate hydratase, is mutated in some tumors. Romero et al. predicted approximately over 130 predicted metabolic pathways in humans [Romero, et al., 2004].

The reaction flux (with initial conditions from Wu et al. (2007)) multiplied by the enzymeÕs half-life (data obtained from experiments by Patrick et al. (2012), the sum (over relevant genes) of 2x (with x the normalized microarray gene expression value obtained from The Cancer Genome Atlas glioblastoma multiforme Agilent dataset) multiplied by the geneÕs half-life (data obtained from experiments by Friedel et al. (2009)). The edge weight is the probability obtained from dividing this edge metabolic activity score by the sum of all edge metabolic activity scores. Preferred additional data used to compute the edge weight can include biological data derived from the cancer patient, involving protein array values to estimate an enzymeÕs translation score (that multiples the enzymeÕs half-life) and C13 data values for estimating the flux of the metabolic reactions.

\section{{\bf{CONCLUSIONS AND FUTURE WORK}}} 

Future implementations of the visualization will annotate the clinicianÕs selected list of sub-paths with a candidate list of drugs, composed of one or more drugs known to disrupt the sub-path or novel drugs. The list for each sub-path will contain at least novel drugs, because there does not yet exist a drug available to knock out every reaction in the human metabolic pathways. These novel drugs are enzyme-inhibitor drugs specific to each metabolic reaction and will be computationally designed (and ideally pre-tested on cancer tissue cultures or in clinical trials), using already developed computational tools: There is currently available open-source virtual screening software to discover candidates of enzyme inhibitors. Auto Dock, http://autodock.scripps.edu, is the open source standard for molecular docking investigations. Open source PyRx, http://pyrx.source forge.net, uses AutoDock to accomplish high-throughput virtual screening. PyRx will be used with the selected enzyme or enzymes as input and output candidates for small molecule enzyme inhibitors. 

Future implementations of the visualization will include animation to track a patientÕs progress during drug therapy. The goal of the animation is to see the rate of change between the previous time step (time t0 of no drug application or time ti of a drug application) and the latest drug application (ti+1) in order to decide if the drug is significantly affecting the desired sub-paths and whether new sub-paths are among the top-ranked sub-paths, causing a new drug to be used in the drug therapy. 
Additionally, uncertainty in the data will be reflected in future implementations of the visualization.

Future work includes evaluating metric accuracy:
Future work will evaluate metric that predicts drug efficacy pre-drug therapy. Possible techniques include using experimental data to estimate accuracy of metric, simulating data and providing error bounds, finding a gold standard to evaluate the metric. Experimental data requirements are that the overall survival time for each cancer patient must be recorded, the type of enzyme-inhibitor drug used in the clinical trial or study must be known, and the gene expression data for each patient in study must be provided. 

To conclude, a web tool was created to guide the clinician in providing a cancer patient with a personalized drug therapy, directly using the patientÕs biological data, as well as biological and cancer data from a variety of prior studies. The philosophy behind the web tool relies on the view that cancer is a metabolic disease and disrupting key metabolic sites in the cancer tissues can best restore a patientÕs health. The current implementation features a graph representing an important metabolic pathway, along with a ranking of top metabolically active sub-paths in this pathway, and provides scores used to rank these sub-paths, along with an additional score that predicts the efficacy of a drug targeting metabolic sites in that patient pre-drug therapy application.

\section{{\bf{ ACKNOWLEDGEMENTS}}}
E. Velasquez wishes to thank the Departamento de Matem\'aticas, Facultad de Ciencias, Universidad de Chile, Santiago for their kind hospitality, and to thank J. Soto-Andrade for hosting her visit and for stimulating conversations. The work conducted by E. Velasquez for this project report was funded by Fondecyt Project No. 1120571.

\section{{\bf{ONCOMIRA TEAM}}}
Elinor Velasquez directed the team, designed the edge weight equation, the algorithm for computing the most likely metabolically active sub-path, and the metric predicting the proposed drug efficacy, collected and pre-processed the data, computed the edge weights, the sub-path ranks, the sub-path metric scores, and also co-designed the website layout and wrote the paper. Ben Bongalon co-designed the website layout, implemented the website, and wrote the description of the code used in the website implementation.

\section{{\bf{REFERENCES}}}

Parsons, DW, Jones, S, Zhang, X, Lin, JC, Leary, RJ, Angenendt, P, et al., An integrated 
genomic analysis of human glioblastoma multiforme. Science 321: 1807Ð1812 (2008). 

Yan, H, Parsons, DW, Jin, G, McLendon, R, Rasheed, BA, Yuan, W, et al., IDH1 and IDH2 mutations in gliomas. New England J Medicine 360: 765Ð773 (2009).

Romero, P, Wagg, J, Green, ML, Kaiser, D, Krummenacker, M, Karp, PD., Computational prediction of human metabolic pathways from the complete human genome. Genome Biol. 6(1): R2 (2005). 

Thiele, et al., A community-driven global reconstruction of human metabolism. Nature Biotechnol. 31(5): 419-25 (2013). 

Karr, JR, Sanghvi, JC, Macklin, DN, Gutschow, MV, Jacobs, JM, Bolival, B, Assad-Garcia, N, Glass, JI, and Covert, MW, A whole-cell computational model predicts phenotype from genotype. Cell 150: 389Ð401 (2012). 

Fan Wu, Feng Yang, Kalyan C. Vinnakota and Daniel A. Beard, Metabolism and Bioenergetics: Computer Modeling of Mitochondrial Tricarboxylic Acid Cycle, Oxidative Phosphorylation, Metabolite Transport, and Electrophysiology. J. Biol. Chem. 282: 24525-24537 (2007). 

Ralph Patrick, Kim-Anh LeCao, Melissa Davis, Bostjan Kobe and Mikael Boden, Mapping the stabilome: a novel computational method for classifying metabolic protein stability. BMC Systems Biology 6:60 (2012). 

Caroline C. Friedel, Lars Dšlken, Zsolt Ruzsics, Ulrich H. Koszinowski and Ralf Zimmer, Conserved principles of mammalian transcriptional regulation revealed by RNA half-life. Nucleic Acids Research 37(17): e115 (2009).

\end{document}